\documentclass[aps,prl,twocolumn]{revtex4}

\usepackage{graphicx}

\begin{document}

\title{Phonon softening and dispersion in the
1D Holstein model of spinless fermions}

\author{C.~E.~Creffield$^1$, G.~Sangiovanni$^2$, and M.~Capone$^{2,3}$}
\affiliation{$^1$ Dipartimento di Fisica, Universit\`a di Roma
``La Sapienza'', Piazzale Aldo Moro 2, I-00185 Rome, Italy}
\affiliation{$^2$ INFM-SMC, Piazzale Aldo Moro 2, I-00185 Rome, Italy}

\affiliation{$^3$ Istituto dei Sistemi Complessi del CNR,
Via dei Taurini 19, I-00185, Rome, Italy}

\date{\today}

\begin{abstract}
We investigate the effect of electron-phonon interaction on the
phononic properties
in the one-dimensional half-filled Holstein model of spinless 
fermions. By means of determinantal Quantum Monte Carlo
simulation we show that the behavior of the phonon dynamics
gives a clear signal of the transition to a charge-ordered phase, 
and the phase diagram obtained in this way is
in excellent agreement with previous DMRG results.
By analyzing the phonon
propagator we extract the renormalized phonon frequency, 
and study how it first softens as the transition is approached and
then subsequently hardens in the charge-ordered phase.
We then show how anharmonic features develop in the phonon propagator,
and how the interaction induces a
sizable dispersion of the dressed phonon
in the non-adiabatic regime.
\end{abstract}

\maketitle

\section{Introduction}

Despite enormous theoretical efforts, effects arising from the
electron-phonon interaction in strongly correlated many-body systems
still remain incompletely understood. Achieving an understanding
of the complicated interplay between electrons and phonons
is, however, essential to explain such diverse phenomena as colossal
magnetoresistance in the manganites\cite{manga}, the Peierls instability in
quasi-1D materials\cite{quasi1d}, and high temperature superconductivity in
alkali-metal doped fullerenes\cite{gunnarssonrev} 
and cuprate compounds\cite{cuprati}.
Even in moderately or weakly correlated materials, the electron-phonon
interaction can give rise to interesting effects which are not
understandable in the framework of the standard theories of
electron-phonon interaction, namely the Migdal-Eliashberg theory of
superconductivity and the Born-Oppenheimer adiabatic principle.
As a notable example, we mention the anomalies in the phononic
properties of superconducting MgB$_2$\cite{mgb2}.

\begin{center}
\begin{figure}
\includegraphics[width=.45\textwidth,clip=true]{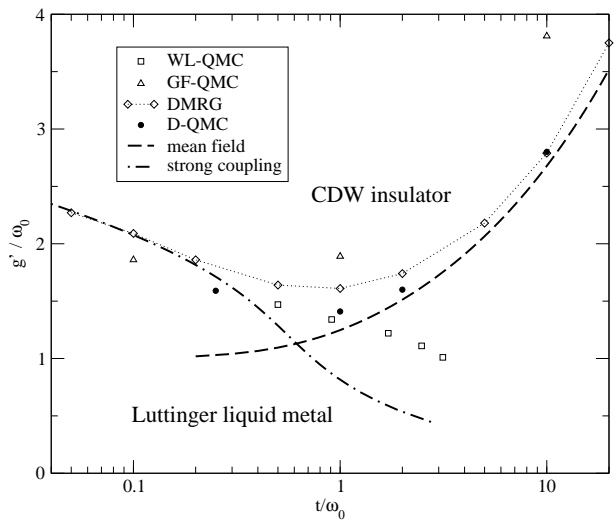}
\caption{Phase diagram of the 1D Holstein model, 
showing the phase boundary between the CDW insulator
and Luttinger liquid metal.
For ease of comparison with previous results, the y-axis
is in units of $g'=g/\sqrt{2 m \omega}$,
the coupling between the electron and the {\em quantized}
phonon operators $a_i / a_i^{\dagger}$.
Squares indicate results from the world-line QMC method from 
Ref.\cite{hirsch}, triangles are results from the Green's function QMC 
investigation of Ref.\cite{mckenzie} and the solid circles are the results
of this work using determinantal QMC. Diamonds denote the results
of the DMRG investigation in Ref.\cite{bursill}.}
\label{phasediag}
\end{figure}
\end{center}

In this work we present a numerical investigation of a simple model
for coupled electron-phonon systems --- the half-filled Holstein 
model for spinless fermions --- and mainly focus on the dynamics of
the phononic degrees of freedom. 
In spite of its apparent simplicity, the model fully
accounts for the competition between local quantum fluctuations and
the tendency for charge ordering, and is thus a powerful tool to 
understand the physics of more realistic systems. 
In one-dimension the system forms a metallic Luttinger liquid 
for weak electron-phonon coupling. As the coupling
is increased, a quantum phase transition occurs
to an insulating state with long-range charge-density-wave (CDW) order.
A wide variety of methods have been used
to investigate this phase transition, with varying degrees of success.
Exact diagonalization schemes \cite{massimo} permit extremely
accurate calculation of the ground-state and low-lying excitations,
but are limited to treating rather small clusters due to the large
Hilbert space required for the phonon degrees of freedom, 
and are thus subject to large finite-size effects.
Larger systems are accessible using the density matrix renormalization
group (DMRG) method, and
in a recent work \cite{bursill} Bursill {\it et al } were
able to determine 
the location of this phase boundary using this approach.
Their result correctly recovers the adiabatic
and anti-adiabatic limits (see Fig.\ref{phasediag}) which can be evaluated
analytically \cite{hirsch}, and 
interpolates smoothly between them. Results obtained
from a variational Lanczos scheme on small clusters \cite{weisse}
agree well with the DMRG result, but data produced by Quantum Monte Carlo
(QMC) methods, such as the worldline QMC method used in 
the pioneering investigation of Ref.\cite{hirsch}
and Green's function QMC \cite{mckenzie}, show significant deviations, the
cause of which is not known.
The role of quantum lattice fluctuations on observables has been discussed in 
Refs. \cite{cinesi,napoletani}, while the spectral properties have been
analysed in Refs. \cite{sykora,hohenadler}

In this work we make use of the determinantal QMC method \cite{blank}
which allows us to compute dynamical properties like the electron and 
phonon propagators without approximations.
These quantities are not equally accessible to DMRG, which is otherwise
the best method to establish the phase diagram of one-dimensional models.
In contrast to previous QMC investigations we are able to obtain
estimates for the location for the phase boundary
in excellent agreement with the ``benchmark'' DMRG result.
We concentrate on the phonon propagator, which appears
to produce more robust results than the electronic properties
considered in previous QMC studies, and show 
firstly how its behavior reveals the location of the phase-boundary.
We then go on to investigate in detail the phonon-softening and 
anharmonicity effects induced by the electron-phonon interaction
by combining analysis of the imaginary-time data 
with the spectral functions obtained by analytically continuing 
this data to real frequencies.
This allows us to identify a renormalized phonon frequency,
and to follow its behavior as a function of electron-phonon coupling.
Beside a generic softening of the frequency, we observe how
for large bare phonon frequency, the harmonic approximation for
the dressed phonon propagator fails as soon as the coupling with
the electrons is introduced. We also discuss how the renormalization of the
phonon properties also gives rise to a phonon dispersion --- even if the
bare phonons are dispersionless.

The paper is organized as follows. In Sec. II we introduce the model and
some details about the QMC simulations. Sec. III presents the results,
and it is in turn divided into two subsections --- the 
phonon propagator in imaginary time and its analytical continuation
to real frequencies. In Sec. IV we summarize and give our conclusions.

\section{Model and simulation}

We consider the following Hamiltonian, describing spinless fermions
moving on a periodic 1D lattice and interacting with a dispersionless
phonon at each lattice site:
\begin{eqnarray}
H = &-&t \sum_{i} \left(  c_i^{\dagger} c_{i+1}^{ } + H.c. \right)
-g \sum_i n_i q_i + \mu \sum_i n_i \nonumber \\
{ } &+& \sum_i \left( \frac{p_i^2}{2 m} + \frac{m \omega_0^2 q_i^2}{2} \right) .
\label{ham}
\end{eqnarray}
Here $c_i^{ } / c_i^{\dagger}$ are the fermion annihilation/creation
operators and $n_i$ is the fermion number operator
$c_i^{\dagger} c_i^{ }$. The electron-phonon coupling is set by $g$,
the phonon frequency is given by $\omega_0$,
and $q_i$ and $p_i$ denote the phonon displacement and momentum operators
respectively. We will express all energies in units of the electronic
hopping $t$, and set the phonon mass $m$ equal to one. 
Unlike Ref.\cite{hirsch} we work in the grand-canonical ensemble, in
which the electronic density is regulated by the chemical potential
$\mu$. In this work we only consider the case of the half-filled system,
which is given by  $\mu = g^2/ 2 \omega_0^2$.

To simulate this model we employ the well-known determinantal QMC
method (DQMC) developed by Blankenbecler, Scalapino and
Sugar \cite{blank}. In this approach the fermion degrees of freedom
are analytically integrated out of the action, which is straightforward
as the Hamiltonian (\ref{ham}) is bilinear in fermion operators,
leaving an {\em effective action} expressed just in terms of the phonon
displacement field. By formally replacing the time coordinate with
imaginary-time ($\tau = it$), the partition function of the model
can then be simulated as a path integral of a Euclidean field theory using
standard Monte Carlo techniques. In this formalism the $\tau$-axis
represents an additional compact dimension, the extent of which
is given by the inverse temperature $\beta$.
In order to sample the zero-temperature properties of the system, it
is important that sufficiently large values of $\beta$ are used.
By comparing the convergence of simulations as $\beta$ was increased
we established that in the adiabatic
regime ($\omega_0 \leq t$) it was sufficient to take $\beta t = 8$, but for
the highest phonon frequency that we studied ($\omega_0=4t$)
a larger value of $\beta t = 16$ was required. Increasing the size of
$\beta$ also requires increasing the number of time-slices used in the
simulation, in order to keep the systematic error arising from the Trotter
decomposition sufficiently small. This both increases the
simulation's running time and also diminishes its numerical stability,
and therefore sets an upper limit on the value of $\omega_0$ we are able
to treat with this method.

Observables, such as the phonon correlation function, are
obtained from the simulation as thermal averages of the form:
\begin{eqnarray}
D_{ij}(\tau) &=& \langle q_i(\tau) q_j(0) \rangle,
		    \quad 0 \leq \tau < \beta  \nonumber \\
	 &=& \mbox{Tr}\left[ q_i(\tau) q_j(0) e^{-\beta H} \right] / Z
\label{propagator}
\end{eqnarray}
where $Z = \mbox{Tr}[\exp(- \beta H)]$, and time-dependent operators are
defined as $q(\tau) = \exp(- H \tau) q \exp(H \tau)$. It is important
to note that the correlation functions produced by the simulation
depend on the imaginary, or Matsubara, time-coordinate $\tau$. To study
the system's dynamical properties it is thus necessary to make an analytic
continuation of these functions to the real-time domain. This amounts
to the solution of the following inverse problem \cite{silver}:
\begin{equation}
S(\tau) = \int_{- \infty}^{\infty} d \omega \frac{ e^{-\omega \tau} }
{1 + e^{-\beta \omega}} \chi_{T}''(\omega) ,
\label{transform}
\end{equation}
where $S(\tau)$ is the Matsubara correlation function, and
$\chi_T'' (\omega)$ is the imaginary component of the time-ordered
susceptibility. To perform this inversion we use a 
maximum entropy technique \cite{silver,bryan}
based on the singular-value decomposition \cite{svd} of the kernel of
Eq.\ref{transform}. By imposing positivity and smoothness constraints on
the solution we find that the
instabilities typically associated with numerical analytical continuation
procedures can be controlled, and this method is thus able to
produce stable results of high resolution for the phonon spectral
function.

\section{Results}

\subsection{The Matsubara phonon propagator}

In principle, the transition from the Luttinger liquid phase
to the CDW insulator can be observed in the staggered-phonon order
parameter
\begin{equation}
m_p = \frac{1}{N} \sum_j (-1)^j \ \langle q_j \rangle ,
\label{mp_def}
\end{equation}
which takes a non-zero value in the CDW phase. It is problematic, however,
to measure this quantity directly in
a simulation of a finite system, as in the CDW phase the system continually 
tunnels between the two degenerate ground-states, causing the
expectation value of $m_p$ to vanish in the ergodic limit.
Over {\em short} runs, however, a non-zero value can be obtained for 
$m_p$ if the system remains trapped in one of the minima for the duration 
of the simulation. This allows the location of the phase transition to be 
located approximately \cite{hirsch}, but
this estimate is intrinsically rough, and cannot be improved easily
as the effect of increasing the number of measurements is to make
$m_p$ vanish. Nonetheless it provides a useful initial comparison for other
estimates, and we found that in all cases it was consistent with
the values we obtained by other methods.

As mentioned above, our interest is devoted to the phonon dynamics, 
which provides us with an alternative way to pinpoint the
phase transition.                 
More precisely, we study the phonon propagator            
(\ref{propagator}), which in the non-interacting case ($g=0$) reads:
\begin{equation}
D_{ij}^0(\tau,\omega_0) = \frac{1}{2 \omega_0}
\frac{ \cosh \omega_0 (\tau - \beta/2)}{\sinh \omega_0 \beta /2} \ \delta_{ij}.
\label{nonintprop}
\end{equation}
It should be noted that as the Holstein phonon is purely local,
the bare phonon propagator also only has a trivial spatial dependence. 
For small electron-phonon coupling it is reasonable to expect
that the phonon propagator has the same
form, but with with a renormalized (softened) frequency $\Omega$ given
by $\Omega^2/\omega_0^2 = 1 - \Pi(0)/\omega_0^2$, where $\Pi(\omega)$ is
the local phonon self-energy in real frequencies.
This ansatz is most accurate when the frequency-dependence
of the phonon self-energy is weak,
and deviations from this ideal behavior
are the fingerprints of phonon anharmonicity, i.e. of the difficulty in 
describing the fully dressed phonon as a single renormalized oscillator.

As the system approaches the CDW transition, the staggered ($k=\pi$)
phonon correlation function is expected to soften and eventually produce
a sharp peak at zero frequency, the weight of which measures the condensation
of the phonons. For an infinite system at $T=0$, this quantity would be the 
phonon staggered-order parameter $m_p$. 
On the other hand, the appearance of zero-frequency weight 
may also occur in the local ($k=0$) correlation function, which
is not, however, directly associated with the CDW transition. 
In this case, the shift of weight can be associated
with a polaron crossover from a good metal to a bad metal, in which the
electronic mobility is strongly reduced by the large coupling
to the phonons.

\begin{center}
\begin{figure}
\includegraphics[width=.45\textwidth,clip=true]{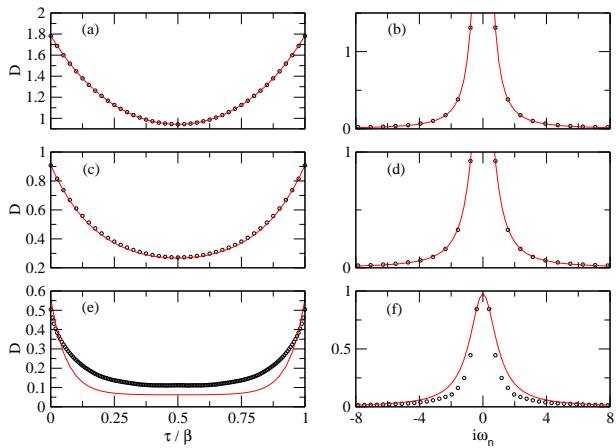}
\caption{(a) The staggered phonon propagator as a
function of imaginary time for $\omega_0 = 0.5$, $g = 0.6$
(c) for $\omega_0 = 1$ and $g = 1.5$
and (e) for $\omega_0 = 4$ and $g = 14.0$.
All data is taken from simulations of 16-site systems.
The line shows the fit to the form
$D(\tau, \omega_0) = D^0(\tau, \Omega) + m_p^2$.
In (b), (d) and (f) we compare the corresponding Fourier transforms 
of these propagators in Matsubara frequency space
with the renormalized Lorentzian Eq.\ref{lorentz}.}
\label{propagators}
\end{figure}
\end{center}

\begin{center}
\begin{figure}
\includegraphics[width=.45\textwidth,clip=true]{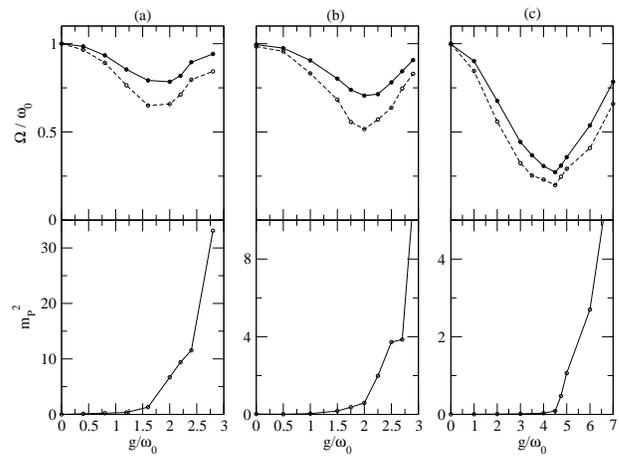}
\caption{Softened  phonon frequency, $\Omega$, and 
order parameter, $m_p^2$, obtained by fitting
the phonon propagator to the form 
$D(\tau) = D^0(\tau,\Omega) + m_p^2$. 
Solid lines indicate the local phonon frequency, dotted lines
the staggered phonon.
(a) $\omega_0 = 0.5$, (b) $\omega_0 = 1$ and (c) $\omega_0 = 4$.
In all cases the staggered phonon is softened more than the local phonon,
and it can be clearly seen that the minimum becomes increasingly deep
(i.e. the phonon softens more) as $\omega_0$ is increased.}
\label{phonon}
\end{figure}
\end{center}

We tested this expectation by fitting the
local and staggered imaginary time correlation functions with the function 
$D^0(\tau,\Omega) + c$, using the softened
frequency $\Omega$ and the shift $c$ as fitting parameters.
The rigid shift in imaginary time $c$ is associated with a $\delta$-like
peak at zero frequency, and describes the static average of the phonon field.
For the {\em local} correlator the value of $c$ indicates the static uniform distortion, which reduces the electron mobility through polaronic effects, 
while for the {\em staggered} case it is simply equal to the staggered 
phonon order parameter $m_p^2$.
For a small value of the bare phonon-frequency $\omega_0=0.5$, we 
find that using the above form for the fitting function 
yields excellent results for both 
the local and staggered propagators, at all values of
the electron-phonon coupling. We show one such example for the staggered
propagator in Fig.\ref{propagators}a.
The good quality of the fit is corroborated by the form
of the Fourier transform of the correlator, shown to the
right. The Fourier representation of the non-interacting
propagator (\ref{nonintprop}) can be easily shown to be
\begin{equation}
D^0_{i j} (i \omega_n, \omega_0) = \frac{1}{\omega_0^2 + \omega_n^2}
\label{lorentz}
\end{equation}
where $\omega_n$ are the Matsubara frequencies,
$\omega_n = 2 \pi n / \beta$. 
Simply replacing the bare frequency, $\omega_0$, in Eq.\ref{lorentz}
with the renormalized frequency, $\Omega$, obtained from the
fitting procedure provides an extremely
good fit to the Fourier transform of the Matsubara
data, as can be seen in Fig.\ref{propagators}b. 
A similar behavior is observed for $\omega_0 = 1$ 
(Figs.\ref{phonon}c,d),
where the fitting procedure again yields excellent results
for both the Matsubara-time propagator and its Fourier transform. 
We emphasize that these fits, which essentially pass through every 
data-point, are obtained using merely {\em two} fitting parameters
with a direct and suggestive physical meaning.

On increasing the bare phonon frequency further, however,
we find that the fitting function no longer
reproduces the data once the
electron-phonon coupling is turned on. 
As can be seen from Fig.\ref{propagators}e the curvature of
the phonon propagator as a function of $\tau$ cannot be described
simply in terms of a softened frequency. This effect can be
seen equally distinctly in frequency space in Fig.\ref{propagators}e,
where $D(i\omega_n)$ clearly deviates from the predicted Lorentzian behavior.
Effectively the non-adiabatic bare phonon introduces anharmonic effects, 
which are enhanced by increasing the coupling. The relationship between 
non-adiabaticity and anharmonicity has been previously proposed for 
MgB$_2$\cite{baffetto}.

In Fig.\ref{phonon} we show the values for the renormalized frequencies,
obtained by the fitting procedure,
for both the local and staggered phonon propagators.
The values of $m_p^2$ produced
by the fitting method are also shown below.
For $\omega_0=0.5$ (Fig.\ref{phonon}a) it can be seen that in the 
metallic region the local
phonon frequencies smoothly reduce from their bare values as $g$
increases, to reach a minimum value at the point $g / \omega_0 = 1.6$.
Throughout this process the staggered phonon-frequency is
softened to a greater degree than the local phonon.
At the local minimum, $m_p$ suddenly acquires
a non-zero value, indicating that this point
signals a transition to the CDW regime, and thus most of the
weight in the phonon propagator moves to the zero frequency peak. 
From Fig.\ref{phasediag}
it can be clearly seen that this estimate compares extremely
well with the location of the phase transition found in Ref.\cite{bursill}.
On increasing $g$ further, the renormalized
frequency is seen to harden \cite{kabanov}, and eventually approach the
bare frequency in the atomic limit, where the system is no longer
metallic and therefore there is no screening of the phonons.

For higher phonon-frequencies a similar behavior
occurs, with the softened phonon frequency passing through
a well-defined minimum at which $m_p^2$ turns on. 
For $\omega_0 = 1$ (Fig.\ref{phonon}b)
the softening of the phonon frequency
is more pronounced, with the staggered phonon frequency decreasing by a factor
of one-half as compared with a factor of about two-thirds for
the previous case. This trend continues for the
antiadiabatic case, $\omega_0=4$ (Fig.\ref{phonon}c),
for which the bare phonon frequency is reduced by almost a factor of ten.
Since the fitting procedure
is less trustworthy for this case, the values obtained for the fitting
parameters should be treated more cautiously. 
The good agreement with the DMRG result that is nonetheless
obtained indicates that the procedure is successfully
describing the phonon dynamics with reasonable accuracy,
and underlines the reliability of using this behavior 
as a signal of the phase transition.

\begin{center}
\begin{figure}
\includegraphics[width=.45\textwidth,clip=true]{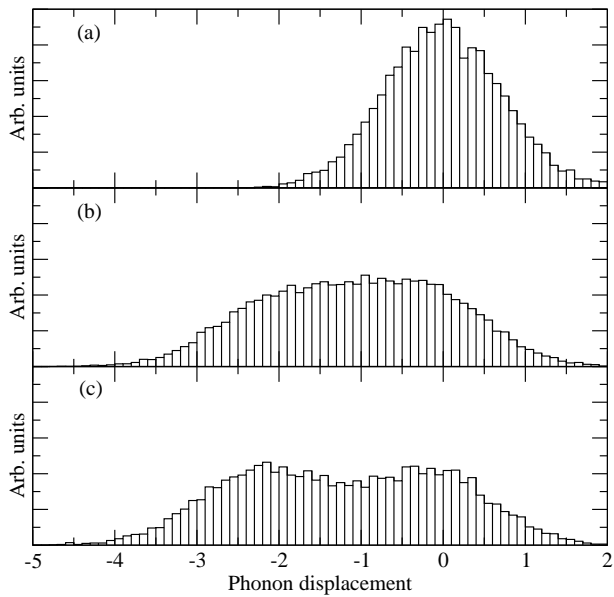}
\caption{ Histograms of the phonon displacement field $\langle q_i \rangle$
in the vicinity of the phase transition for a system
with phonon frequency $\omega_0 = 1$. (a) $g=0.0$, (b) $g=2.25$
and (c) $g=2.5$. Note how the non-interacting
Gaussian distribution centered on zero
first flattens as the phase transition is approached,
and then splits into two in the CDW regime,
indicating the formation of a polaronic state.}
\label{histogram}
\end{figure}
\end{center}

Further insight about the effect of the interaction on phononic
properties can be gained by the analysis of the statistical distribution
of the phonon displacement field $q_j$. In Fig. \ref{histogram} we show how
this quantity evolves as the electron-phonon coupling is increased.
A clear qualitative change occurs from an unimodal distribution in the
metallic phase, in which $q=0$ is the most probable value, to a bimodal
distribution with maxima \cite{shift}
separated by $2 q_0 = g / \omega_0^2$,
which signals the formation of local lattice distortions
peaked around $\pm q_0$.
In previous work \cite{whatever}, the evolution of the distribution of the
displacements has been used to characterize the polaron crossover 
in the Holstein model. In principle, the formation of local 
distortions does not automatically imply a CDW ordering, as has been 
discussed for example in Ref.\cite{caponeciuchi}. For the system we study, 
however, it does appear that the formation of local lattice distortions and
CDW ordering do occur simultaneously, implying that a polaronic
metal state is not present in the transition region.

\subsection{Analytic continuation of the phonon propagator}

To complement the previous analysis of the Matsubara correlation
functions, we now study the real-frequency propagator by employing
a maximum entropy method to make the analytic continuation
from Matsubara time to real-frequencies.
This avoids the need of assuming a given analytic form for
the propagator, and also allows us
to study the softening of the phonon in more detail.
In Fig.\ref{analcon} we present contour plots of the local phonon
spectral function for different bare phonon-frequencies. Darker areas
correspond to larger weights. In all figures, the weight is clearly
concentrated at the bare frequency ($\pm \omega_0$) at $g=0$, 
and for weak coupling it remains concentrated in a single 
feature at the renormalized frequency $\Omega$. 
As the renormalized phonon softens further,
a zero frequency peak appears and takes most
of the weight. In terms of the fitting parameters, the appearance
of this feature corresponds to a non-zero value of the shift $c$,
implying a sizable static lattice deformation, which reduces
the mobility of the electrons.
After the CDW transition a higher energy phonon branch forms, 
and, as was seen previously from the Matsubara analysis,
its energy hardens and eventually
converges to the bare frequency in the limit of extreme strong-coupling.
In Fig.\ref{analcon} the formation of this branch is clearly
visible for the non-adiabatic cases $\omega_0=1$ and 4, and
although it also occurs in the adiabatic case the higher energy features
are rather obscured by the large zero frequency peak.
Thus, in the limit of strong-coupling the system recovers the appearance
of the atomic limit in
which the phonon is unrenormalized, due to the lack of metallic
screening.

\begin{center}
\begin{figure}
\includegraphics[width=.225\textwidth]{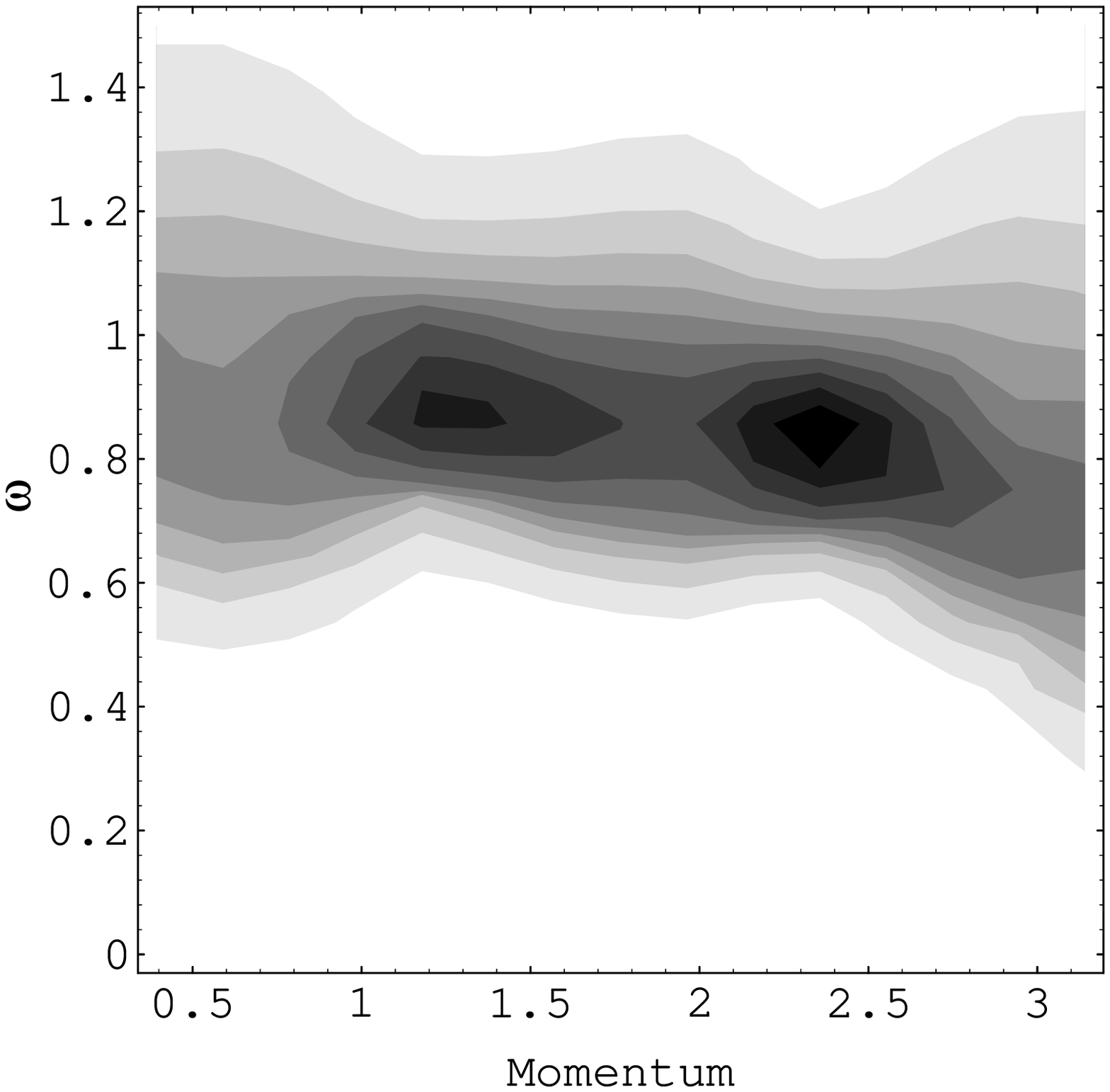}
\includegraphics[width=.225\textwidth]{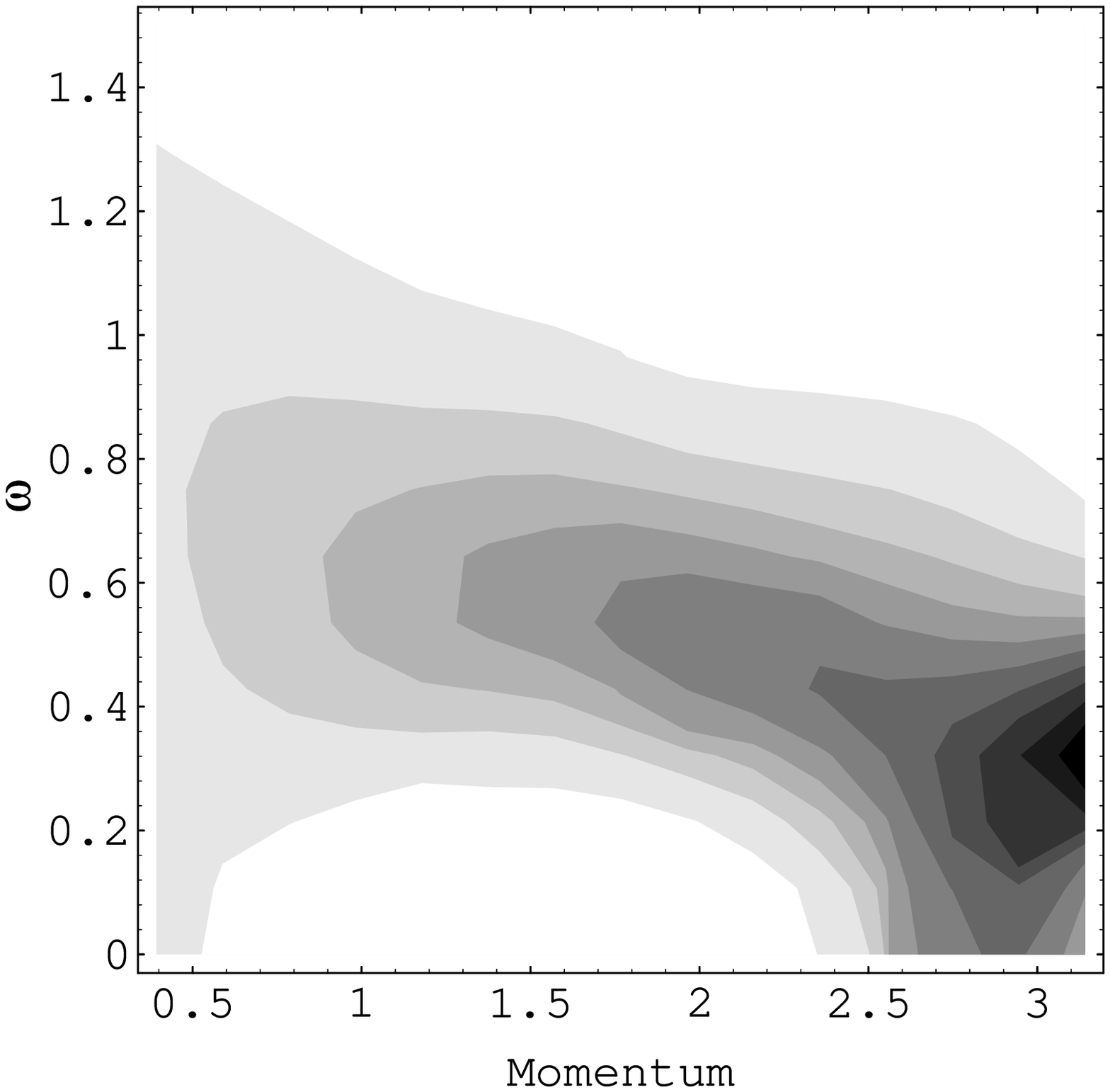}
\caption{Phonon susceptibility $\chi''(k,\omega)$ for
$\omega_0=1$. The peaks in this quantity (dark areas)
indicate the dispersion relation followed by the dressed
phonon. Left: $g=1.0$, the dressed phonon is practically
dispersionless, resembling the bare Holstein phonon.
Right: $g=2.0$, the phonon peak shifts toward zero as 
$k$ increases indicating that the dressed phonon has acquired
a substantial dispersion from the electron-phonon interaction.}
\label{dispersion}
\end{figure}
\end{center}

A similar analysis can be carried out for all the values of the
exchanged phonon momentum $k$, resulting in a momentum dependent
renormalized phonon frequency $\Omega(k)$. We plot this quantity for
$\omega_0=1$ and $g=1$ and $2$ in Fig. \ref{dispersion}.
In the weaker coupling case, the renormalized phonon is still basically
dispersionless like the bare Holstein phonon. For the larger coupling, 
however, the renormalized phonon becomes dispersive, 
exhibiting an approximately $\cos (k/2)$ dependence on momentum.
This dramatically demonstrates that if a material has a sizable electron-phonon
interaction, the bare dispersion of the phonons may be substantially 
different from the fully dressed one. This has to be seriously taken into
account in the derivation of effective models, in which  the 
experimentally observed {\it fully dressed} dispersion should not be used as
a bare dispersion for a model calculation to avoid double counting. 

\begin{widetext}
\begin{center}
\begin{figure*}
\includegraphics[width=.3\textwidth]{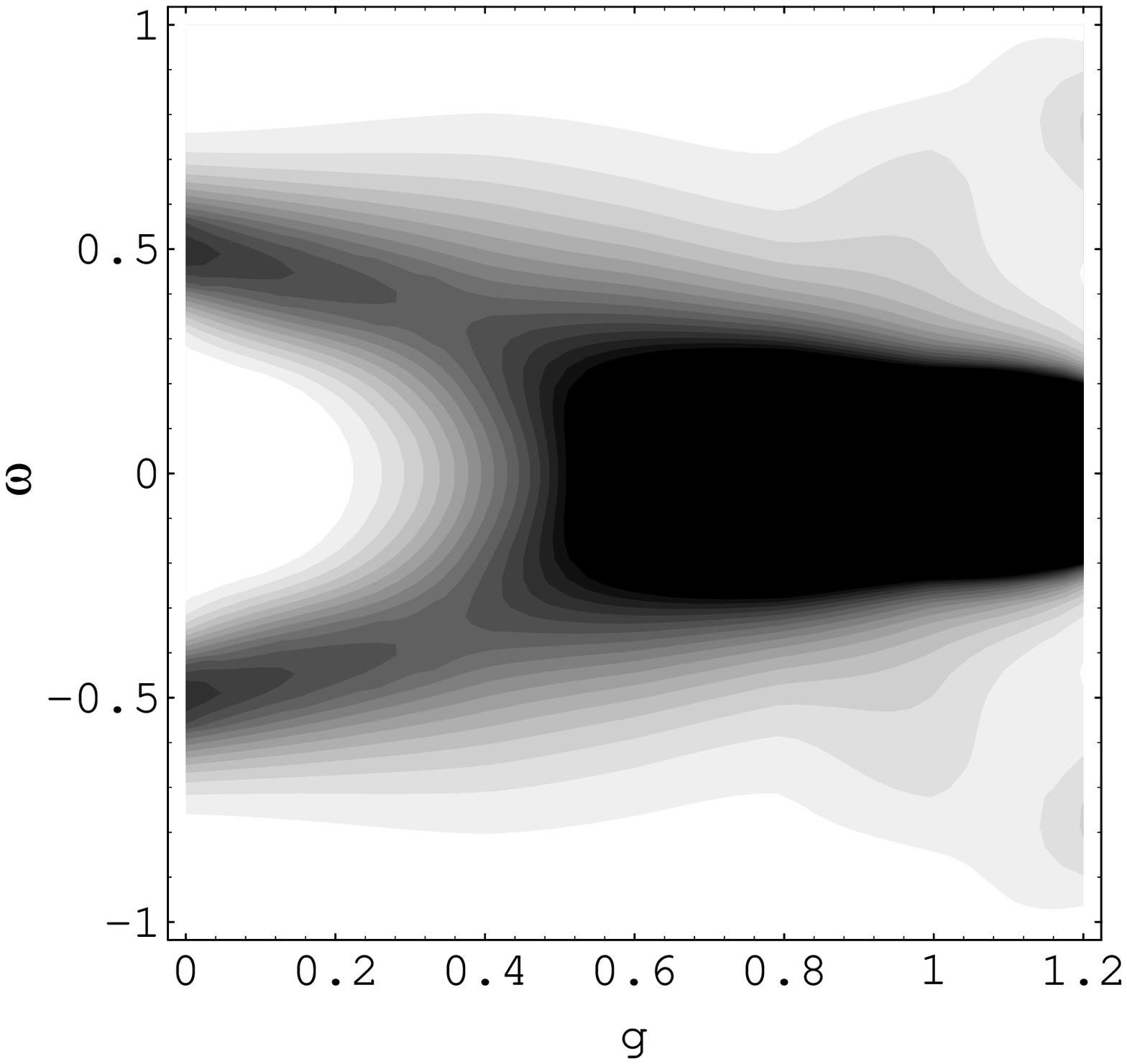}
\includegraphics[width=.3\textwidth]{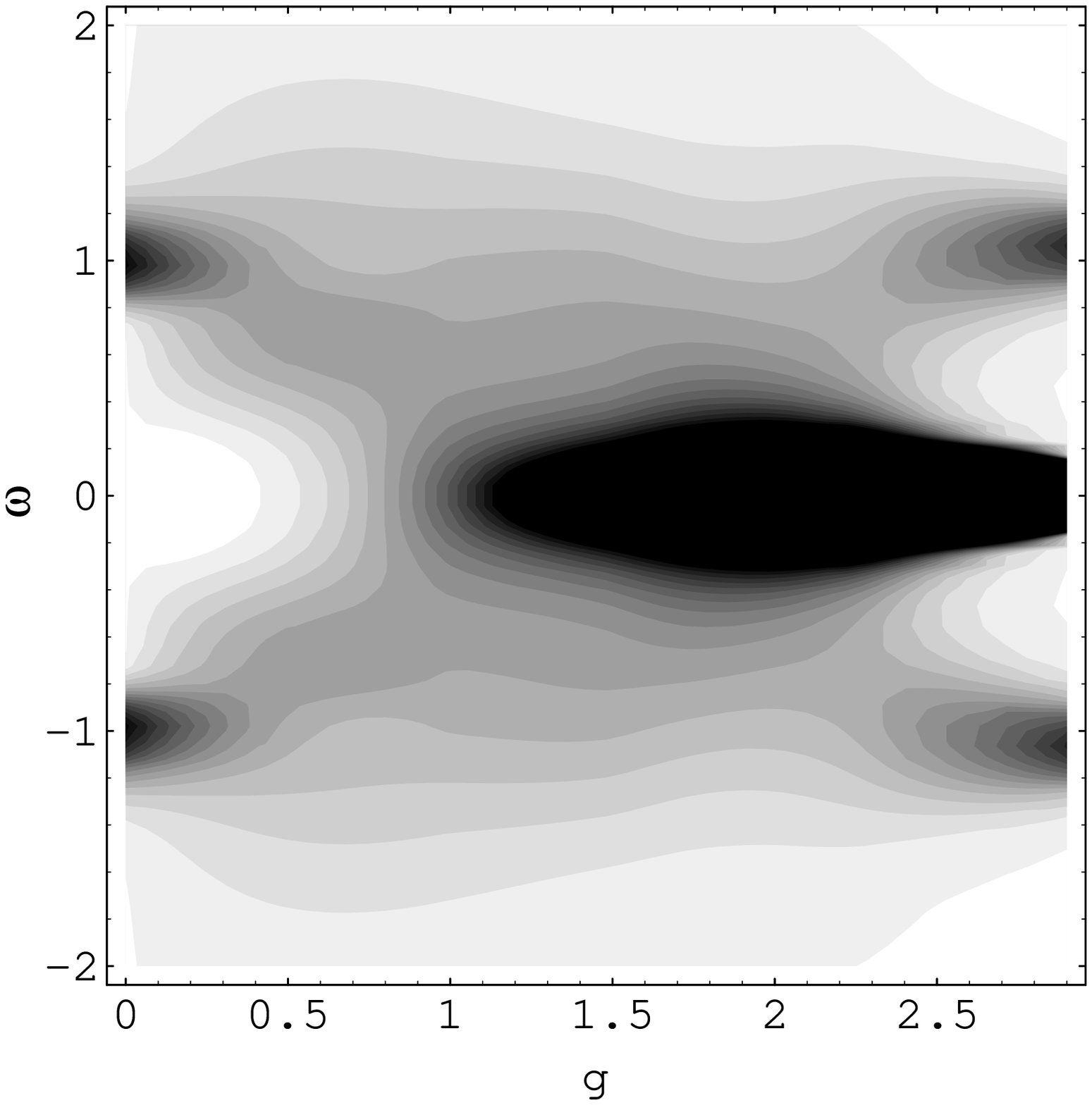}
\includegraphics[width=.3\textwidth]{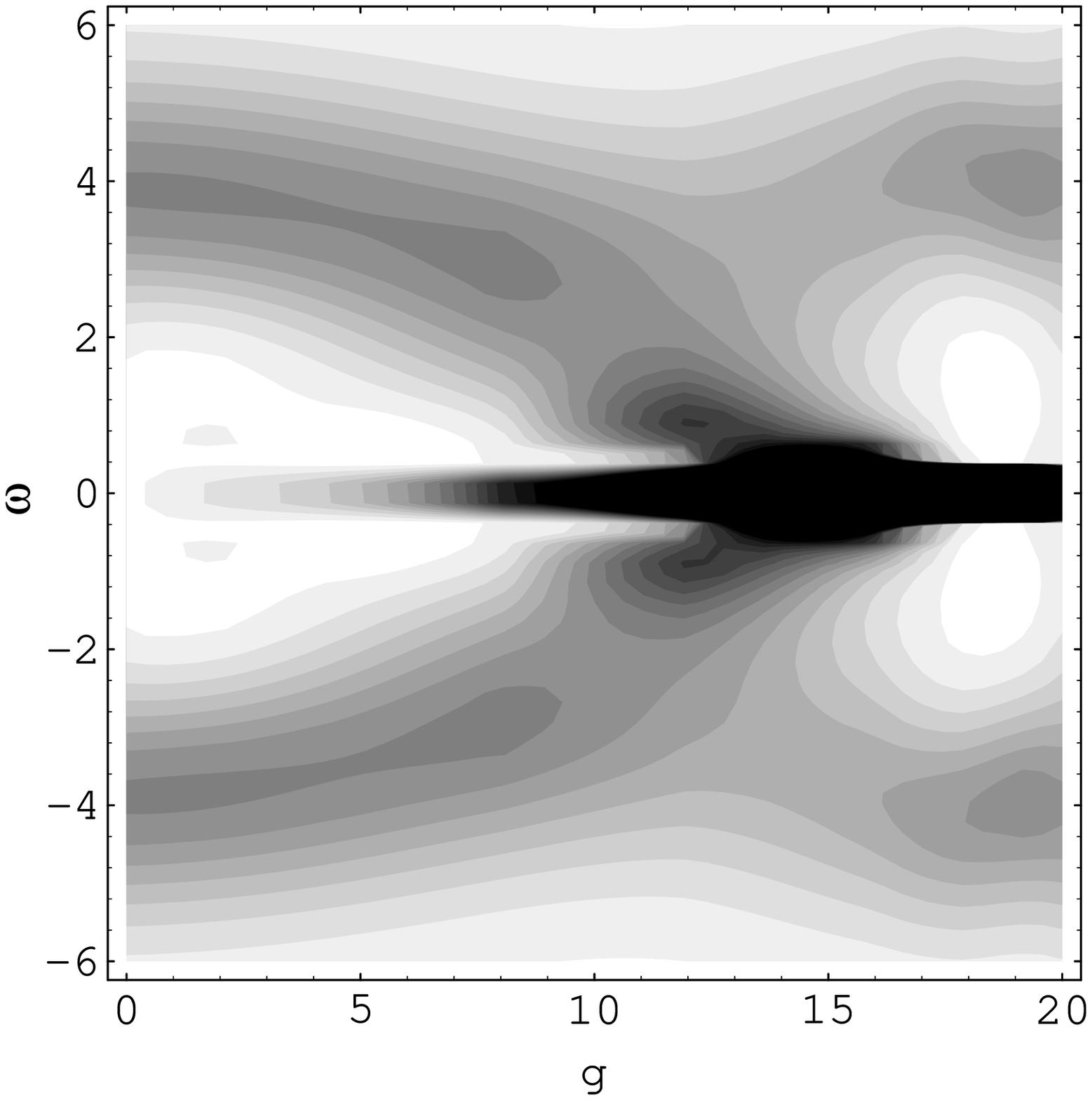}
\caption{Analytic continuation of the local phonon propagator 
$D_{i i}(\tau)$ for phonon frequencies $\omega_0 = 0.5, 1.0, 4.0$.}
\label{analcon}
\end{figure*}
\end{center}
\end{widetext}

\section{Conclusions}

In this paper we have investigated the effect of electron-phonon interaction
on the half-filled one-dimensional Holstein model by means of DQMC 
simulations. We have firstly shown how examination of the phonon 
propagator can be used to determine the boundary for the CDW
phase transition, which agree well with the state-of-the-art
DMRG data. 

The agreement of our results with the DMRG phase diagram makes us 
confident about the reliability of our method in the evaluation of 
other observables, which are more difficult to obtain with DMRG, such as 
dynamical properties. 

In particular, we focus on the phonon propagator, in order to discuss the
dressing of the phonon degrees of freedom for large coupling with the
electrons. 
An analysis of the Matsubara phonon propagator allows 
a renormalized phonon frequency to be deduced directly
from the data produced by the QMC simulation. This has revealed
how the phonon frequency softens as the coupling is increased
until, at a critical value of $g$, the order parameter $m_p^2$
turns on and the system makes a phase transition to the CDW regime.
The degree of softening is considerably higher in the non-adiabatic
regime as compared to the adiabatic case. 
Our analysis has also highlighted the relation between non-adiabatic
effects and anharmonicity. When the bare phonon frequency is small,
the phonon propagator describes basically a single harmonic phonon
even for large coupling, while in the non-adiabatic case,
the electron-phonon coupling induces anharmonic effects through
a frequency dependent phonon self-energy.

By using analytic continuation methods we were able
to study the phonon renormalization in detail.
At the critical point we have seen how
the phonon mode splits into two --- one mode developing into
a soft mode, coexisting with the other mode which then hardens
and re-approaches the bare frequency as $g$ is increased further.
This technique has also revealed the strong
momentum-dependence of the dressed phonon in the non-adiabatic
regime, arising from the electron-phonon interaction.
In particular, a sizable phonon dispersion arises as the 
coupling is increased, even starting from a bare dispersionless
phonon, suggesting that care must be taken in building up
models for electron-phonon interaction starting from experimentally
observed phonon properties. 

\begin{acknowledgments}
We thank C. Castellani and S. Ciuchi for valuable discussions. 
We also acknowledge Italian MIUR Cofin2003 for financial support.
\end{acknowledgments}


\begin{thebibliography}{99}

\bibitem{manga} J. M. De Teresa, M. R. Ibarra, P. A. Algarabel, C. Ritter, 
C. Marquina, J. Blasco, J. Garcia, A. del Moral, and Z. Arnold, 
Nature {\bf 386}, 256 (1997); 
A. J. Millis, {\it ibid.} {\bf 392}, 147 (1998); 
M. B. Salamon and M. Jaime, Rev. Mod. Phys. {\bf 73}, 583 (2001).

\bibitem{quasi1d} {\it Organic Conductors} edited by J.-P. Farges,
(Marcel Dekker, New York, 1994).

\bibitem{gunnarssonrev} M. Matus, H. Kuzmany, and E. Sohmen, 
Phys. Rev. Lett. {\bf 68}, 2822 (1992); 
K. Harigaya, Phys. Rev. B {\bf 45}, 13676 (1992); 
B. Friedman, {\it ibid.} {\bf 45}, 1454 (1992);
W. M. You, C. L.  Wang, F. C. Zhang, and Z. B. Su, {\it ibid.} 
 {\bf 47}, 4765 (1993).

\bibitem{cuprati} Guo-Meng-Zhao, M. B. Hunt, H. Keller, and K. A. Muller,
Nature {\bf 385}, 236 (1997); 
A. Lanzara, P. V. Bogdanov, X. J. Zhou, S. A. Kellar, D. L. Feng, E. D. Lu, 
T. Yoshida, H. Eisaki, A. Fujimori, K. Kishio, J.-I. Shimoyama, T. Noda, 
S. Uchida, Z. Hussain, and Z.-X. Shen, {\it ibid.}
{\bf 412}, 510 (2001); 
R. J. McQueeney, J. L. Sarrao, P. G. Pagliuso, P. W. Stephens, 
and R. Osborn, Phys. Rev. Lett. {\bf 87}, 77001 (2001).

\bibitem{mgb2}
{J. Kortus, I. I. Mazin, K. D. Belashchenko, V. P. Antropov,
and L. L. Boyer, Phys. Rev. Lett. {\bf 86}, 4656 (2001);
P. Postorino, A. Congeduti, P. Dore, A. Nucara, A. Bianconi,
D. Di Castro, S. De Negri, and A. Saccone,
Phys. Rev. B {\bf 65}, 020507R (2002).} 

\bibitem{massimo}
{M. Capone, W. Stephan and M. Grilli, Phys. Rev. B {\bf 56}, 4484 (1997).}

\bibitem{bursill}
{Robert J. Bursill, Ross H. McKenzie and Chris J. Hamer, Phys. Rev. Lett
{\bf 80}, 5607 (1998).}

\bibitem{hirsch}
{Jorge E. Hirsch and Eduardo Fradkin, Phys. Rev. B {\bf27}, 4302 (1983).}

\bibitem{weisse}
{A. Wei\ss e and H. Fehske, Phys. Rev. B {\bf 58}, 13526 (1998).}

\bibitem{mckenzie}
{Ross H. McKenzie, C. J. Hamer and D. W. Murray, Phys. Rev. B{\bf 53},
9676 (1996).}

\bibitem{cinesi}
{Zhiguo L{\" u}, Qin Wang, and Hang Zheng, 
Phys. Rev. B {\bf 69}, 134304 (2004).}   

\bibitem{napoletani}
{C. A. Perroni, V. Cataudella, G. De Filippis, G. Iadonisi, V. Marigliano
Ramaglia, and F. Ventriglia, Phys. Rev. B 67, 214301 (2003).} 

\bibitem{sykora}
{S. Sykora, A. Huebsch, K. W. Becker, G. Wellein, and H. Fehske,
Phys. Rev. B {\bf 71}, 045112 (2005).}

\bibitem{hohenadler}
{Martin Hohenadler, Markus Aichhorn, and Wolfgang von der Linden,
Phys. Rev. B {\bf 68}, 184304 (2003).}
 
\bibitem{blank}
{R. Blankenbecler, D. J. Scalapino and R. L. Sugar, Phys. Rev. D {\bf 24},
2278 (1981).}

\bibitem{silver}
{J. E. Gubernatis, M. Jarrell, R. N. Silver, and D. S. Sivia,
Phys. Rev. B {\bf 44}, 6011 (1991).}

\bibitem{bryan}
{R. K. Bryan, Eur. Bio. J. {\bf 18}, 165 (1990).}

\bibitem{svd}
{C. E. Creffield, E. G. Klepfish, E. R. Pike and S. Sarkar, Phys. Rev. Lett.
{\bf 75}, 517 (1995).}

\bibitem{baffetto}
{L. Boeri, G. B. Bachelet, E. Cappelluti and L. Pietronero, 
Phys. Rev. B {\bf 65}, 214501 (2002).}

\bibitem{kabanov}
{A. S. Alexandrov, V. V. Kabanov and D. K. Ray,
Phys. Rev. B {\bf 49}, 9915 (1994).}

\bibitem{shift}
{As we do not use an explicitly symmetric form for the electron-phonon
interaction term  in Eq.\ref{ham} 
(i.e. we use $g q n$ instead of $g q \left( n - \langle n \rangle \right)$),
this bimodal distribution is not symmetric about the origin, instead having
an overall shift of $-q_0$.}

\bibitem{whatever}
{M. Capone and S. Ciuchi, Phys. Rev. B {\bf 65}, 104409 (2002).}

\bibitem{caponeciuchi}
{M. Capone and S. Ciuchi, Phys. Rev. Lett. {\bf 91}, 186405 (2003).}

\end{thebibliography}
\end{document}